\def\astrobj#1{#1}
\documentclass[]{elsart}
\usepackage{graphicx}
\usepackage{amssymb}
\begin{document}
\begin{frontmatter}

   \title{Searching for radiative pumping lines of OH masers:
         II. The 53.3\,$\mu$m absorption line towards 1612\,MHz OH maser sources.
         \thanksref{ISO}}
	 \thanks[ISO]{Based on observations with ISO, an ESA project with instruments funded by 
		    ESA Member States (especially the PI countries: France, Germany, the 
		    Netherlands and the United Kingdom) and with the participation of ISAS and NASA.}

   \author{J. H. He} and
   \ead{mailhejh@yahoo.com.cn}
   \author{P. S. Chen}

   \address{ ( National Astronomical Observatories/Yunnan Observatory, Chinese Academy of Sciences, Kunming,  PRC )}

   \begin{abstract}

   This is the second paper in a series aiming at searching for infrared pumping lines for galactic 1612\,MHz OH masers. Our paper I is devoted to the 34.6\,$\mu$m absorption lines in ISO SWS spectra towards a large sample of galactic OH/IR sources. This paper analyzes the 53.3\,$\mu$m line in the ISO LWS spectra towards a similar sample of OH/IR sources. A search with position radius of 1 arcmin in ISO Data Archive (IDA) results in 137 LWS spectra covering 53.3\,$\mu$m associated with 47 galactic OH/IR sources and 4 ones associated with megamasers \astrobj{Arp\,220} and \astrobj{NGC\,253}. (These two magamasers are included for comparison purpose only.) Ten of these galactic OH/IR sources are found to show and another 5 ones tentatively show the 53.3\,$\mu$m absorption while another 7 sources (our group U1 and U2 sources) highly probably do not show this line. The source class is found to be correlated with the type of spectral profile: red supergiants (RSGs) and AGB stars tend to show strong {\em blue-shifted} filling emission in their 53.3\,$\mu$m absorption line profiles while H\,II regions tend to show a weak {\em red-shifted} filling emission in the line profile. GC sources and megamasers mainly show symmetrical profile in the line core while megamasers tend to show an additional absorption tail on the blue side of the line profile. It is argued that the filling emission might be the manifestation of an unresolved half emission half absorption profile of the 53.3\,$\mu$m doublet which might be produced by the transitions among the two levels: $^{2}\Pi _{1/2} (J=3/2)$ and $^{2}\Pi _{1/2} (J=5/2)$ and their closely related levels. The 53.3 to 34.6\,$\mu$m equivalent width (EW) ratio is close to unity for RSGs but much larger than unity for GC sources and megamasers while H\,II regions only show the 53.3\,$\mu$m line. The {\em pump rate} defined as maser to IR photon flux ratio is approximately $5\%$ for RSGs. The pump rates of GC sources are three order of magnitude smaller. Both the large 53.3 to 34.6\,$\mu$m EW ratio and the small pump rate of the GC OH masers reflect that the two detected `pumping lines' in these sources are actually of interstellar origin. The pump rate of \astrobj{Arp\,220} is $32\%$---much larger than that of RSGs, which indicates that the contribution of other pumping mechanisms to this megamaser is important. A handful of non--detections of the 34.6 or 53.3\,$\mu$m line or both can be explained partly by the genuinely weakness of the OH masers and partly by some other mechanisms weakening the IR pumping lines, such as clumpy OH shell or limb filling emission.

\end{abstract}



\end{frontmatter}
%

\section{Introduction}

There are mainly three kinds of pumping mechanisms for stable OH maser: by infrared photon, by collision or by chemical reaction(Elitzur,\,\cite{eli92}). Harvey et al.\,(\cite{har74}) favoured the radiative pumping mechanism for 1612\,MHz OH maser in the envelope of OH/IR stars because they found that the OH maser and IR flux were varying together with almost no phase delay in these sources. Elitzur\,(\cite{eli92}) had systematically studied the observational facts and the theoretical considerations of the pumping mechanism of the OH masers. In circumstances where the radiative pumping is favored, the 34.6 and 53.3\,$\mu$m photons had been thought to be the main pumping photons to excite the OH molecules from the ground rotational state $^{2}\Pi_{3/2}(J=3/2)$ to their upper states $^{2}\Pi_{1/2}(J=5/2)$ and $^{2}\Pi_{1/2}(J=3/2)$ respectively. The exited OH molecules then cascade down to invert the $F=1$ and $F=2$ fine-structure levels in the ground state. But astronomers had no chance to probe these absorption lines until the Infrared Space Observaotry (ISO - Kessler et al.\,\cite{kes96}) opened mid-- and  far--infrared (IR) observations otherwise impossible behind the veiling of the atmosphere.
Justtanont et al.\,(\cite{jus96}) and Lutz et al.\,(\cite{lut96}) were the first to find in ISO spectra the 34.6\,$\mu$m absorption line associated with OH maser sources. The former found it in \astrobj{NML\,Cyg}, a Red SuperGiant (RSG), while the latter found it in \astrobj{Sgr\,A*}, the Galactic Center (GC). The next year, Sylvester et al.\,(\cite{syl97}) also profitted from the ISO short wavelength spectrometer and long wavelength spectrometer (SWS \& LWS) observations to find the IR pumping absorption line at 34.6\,$\mu$m together with the OH cascade emission lines at 98.7, 163 and 79$\mu$m towards another well known RSG: \astrobj{IRC+10420}. Thai-Q-Tang et al.\,(\cite{tha98}) modelled the pumping of the OH maser in the envelope of this star. Their work confirmed the radiative pumping mechanism for the stellar 1612\,MHz OH maser. The 34.6\,$\mu$m absorption in another RSG, \astrobj{VY CMa}, was also reported by Neufeld et al.\,(\cite{neu99}). These previous works call for systematic check for infrared pumping lines of the stellar OH 1612\,MHz masers.

In our previous paper (He et al.\,\cite{He04}, here after Paper I), the 34.6\,$\mu$m absorption is found in only 3 RSGs, 2 GC sources  and a megamaser \astrobj{Arp\,220} and tentatively in another two objects (one H\,II region and one Mira) while it is expected but not detected in another 16 galactic OH/IR sources (these are mainly OH/IR stars, Miras and PNe). Several mechanisms suggested for explaining non--detections remain open for further analysis. 

In this paper, the 53.3\,$\mu$m spectrum of 102 (out of 137) ISO LWS spectra associated with 47 galactic OH/IR sources and 4 ISO LWS spectra associated with two megamasers are analyzed. Some new 53.3\,$\mu$m line detection and non--detection sources are found. The 53.3\,$\mu$m line profiles are classified, discussed and correlated to SIMBAD object classes and the detection of 34.6\,$\mu$m line. In Sect.\,\ref{obsprocess}, the observational data is described. The detectability of the 53.3\,$\mu$m line is discussed in Sect.\,\ref{detectability}. The dependence of the OH maser efficiency on the ISO IR color is presented in Sect.\,\ref{ISOIRc}. The profiles are classified in section \,\ref{profiles}. Section \ref{comp35and53} compares the results of the 53.3 and 34.6\,$\mu$m lines while the OH pump rate of several sources are derived in Sect.\,\ref{OHpump}. A brief summary is given in Sect.\,\ref{summary}.

\section{Observation and data processing}
\label{obsprocess}

ISO had observed more than 100 OH/IR sources during its 26 months mission, but not all of them were observed spectroscopically around 53.3\,$\mu$m. A search is performed in the ISO Data Achive (IDA) for associations with OH/IR sources based on a large OH/IR source compilation which combines the catalogue provided by B.M. Lewis (including Eder, Lewis and Terzian\,\cite{ede88}, Lewis, Eder and Terzian\,\cite{lew90}, Lewis\,\cite{lew92}, Chengalur et al.\,\cite{che93} and Lewis\,\cite{lew94}) with OH/IR catalogues from ATCA telescope(Sevenster et al.\,\cite{sev97a}; \cite{sev97b}; \cite{sev01}) and similar catalogues from many other papers (this unpublished large compilation contains 1876 OH/IR sources). Chen et al.\,(\cite{che01}) cross-correlated this large OH/IR compilation with the infrared astronomical satellite low resolution spectral catalogue (IRAS/LRS) and discussed the LRS properties of these sources. An extension of Chen's table of OH/IR sources with LRS identification was also given in electronical form in Paper I. In the present paper, the same OH/IR source compilation is cross-correlated with the ISO/LWS spectra database and 137 ISO/LWS spectra covering the 53.3\,$\mu$m region are found to be associated with 47 galactic OH/IR sources within a positional error of 1$^{\prime}$\,. Note that the IRAS position is used to do the identification, hence the identification results given here can be slightly different from that automatically generated by IDA because the SIMBAD positions of the sources used by IDA are not necessarily the same as their IRAS positions. The details of the cross-correlation are given in Table\,\ref{sptable} in which only the 102 spectra (out of 137) for Galactic OH/IR sources and the 4 spectra of megamasers are listed. The columns in the table are: (1)\,the source name; (2)\,target dedicated time (TDT, the unique name of each ISO spectra); (3)\,ISO LWS spectral resolution expressed in minimum resolved wavelength separation in [$\mu$m]; (4)\,continuum flux at 53.3\,$\mu$m and dispersion (5)  as the mean standard deviation of all flux data points within the $53\sim 54\,\mu$m range. 
%
\begin{table}[]{}
\caption[]{OH/IR sources with their associated ISO LWS spectra covering the 53.3\,$\mu$m region. 
Groups and superscripts are explained in the text. Two megamaser galaxies are included in group `A'.}
\label{sptable}
\begin{tabular}{l@{ }l@{  }l@{  }l@{  }l@{  }}
\hline
\noalign{\smallskip}
      Name             &TDT        &R[$\mu$]    &{\it F}$^{c}_{53.3}$[Jy] &$\sigma$[Jy]  \\
     (1)               &(2)        &(3)         &(4)                    &(5)           \\
\noalign{\smallskip}
\hline
\noalign{\smallskip}
\multicolumn{5}{l}{\textbf{Group A}}                                 \\
\noalign{\smallskip}
$\astrobj{03507+1115}$        &65601707$^a$       &0.07     &427     &2.1 	\\
$\astrobj{06053-0622}$        &71002509$^a$       &0.07     &15912   &33.3 	\\
$\astrobj{07209-2540}$        &73502338$^a$       &0.07     &1890    &3.2 	\\
$\astrobj{17424-2852}$        &32600904$^a$       &0.035    &5182    &8.9 	\\
$\astrobj{17424-2859}$        &49801004$^a$       &0.14     &18284   &145.8	\\
$\astrobj{17430-2848}$        &50701028$^a$       &0.071    &1847    &19.1	\\
$\astrobj{17441-2822}$        &32601008$^a$       &0.035    &6342    &18.9	\\
                              &32701306$^t$       &0.27     &6517    &129.4	\\
                              &49400302$^a$       &0.03     &5775    &31.2	\\
$\astrobj{17574-2403}$        &09901026$^{*a}$    &0.07     &15908   &21.7   	\\
                              &12500924$^t$       &0.14     &14975   &54.4   	\\
$\astrobj{20255+3712}$        &01301603$^a$       &0.07     &9832    &22.6	\\
                              &12600515$^t$       &0.07     &8425    &38.7	\\
                              &13400331$^t$       &0.07     &8462    &55.6	\\
                              &18204615           &0.07     &8837    &147.5 	\\
                              &20306815$^t$       &0.069    &8915    &60.1   	\\
                              &21002015$^t$       &0.071    &8950    &115.9  	\\
                              &22302315$^t$       &0.07     &8778    &48.7	\\
                              &22302801$^a$       &0.071    &8457    &28.3   	\\
                              &35201327$^t$       &0.14     &8103    &78.2	\\
                              &49602708$^t$       &0.069    &8659    &104.9	\\
                              &52403008$^t$       &0.07     &8669    &38.1 	\\
                              &53000508$^t$       &0.27     &8589    &195.6 	\\
                              &53102508$^t$       &0.07     &8534    &37.1	\\
            \noalign{\smallskip}
            \hline
         \end{tabular}
\end{table}
\begin{table}[]{}
\leftline{(Table\,\ref{sptable} continued)}
\begin{tabular}{l@{ }l@{  }l@{  }l@{  }l@{  }}
\hline
\noalign{\smallskip}
     (1)               &(2)        &(3)         &(4)        &(5)   \\
\noalign{\smallskip}
\hline
\noalign{\smallskip}
                              &53803008$^a$       &0.07     &8557    &30.2	\\
                              &54403605$^a$       &0.07     &8603    &36.0 	\\
                              &55205608$^t$       &0.07     &8636    &50.5 	\\
                              &55903008           &0.07     &8600    &27.3	\\
                              &57302606$^t$       &0.07     &8712    &51.6	\\
                              &57902706$^a$       &0.07     &8589    &27.1	\\
                              &58702901$^a$       &0.07     &8703    &28.0	\\
                              &70601601$^t$       &0.07     &8897    &88.3	\\
                              &76902601$^t$       &0.07     &8772    &57.2	\\
                              &77602401$^t$       &0.07     &8972    &49.8	\\
                              &86000201$^t$       &0.071    &8935    &40.4   	\\
$\astrobj{Arp\,220}$          &27800202$^a$       &0.069    &112.8   &1.2 	\\
                              &64000801$^a$       &0.069    &121.2   &2.1 	\\
                              &64000916$^a$       &0.068    &121.2   &2.3 	\\
$\astrobj{NGC\,253}$          &24701103$^a$       &0.069    &964.7   &4.4 	\\
$\astrobj{NML\,Cyg}$          &34201304$^a$       &0.068    &1217    &5.7		\\
                              &55000813$^m$       &0.0015   &1919    &517.2  	\\
                              &55500942$^{ma}$    &0.068    &1178    &3.3  	\\
\noalign{\smallskip}
\multicolumn{5}{l}{\textbf{Group T}}                                 \\
\noalign{\smallskip}
$\astrobj{06319+0415}$        &87102811$^t$       &0.07     &1019    &3.5    \\
$\astrobj{16342-3814}$        &08402827$^t$       &0.071    &346.5   &4.0    \\
$\astrobj{17431-2846}$        &46400917$^t$       &0.14     &3697    &61.0   \\
$\astrobj{18348-0526}$        &33000316$^t$       &0.071    &398     &4.6    \\
                              &34000102           &0.14     &393.1   &4.0    \\
                              &47201318$^m$       &0.07     &412.8   &7.0    \\
$\astrobj{19244+1115}$        &31601203$^t$       &0.071    &795.5   &4.1    \\
                              &36401611$^m$       &0.0015   &255.9   &613.8  \\
                              &72400312           &0.14     &807.5   &4.8    \\
                              &72400414$^t$       &0.14     &794.2   &6.0    \\
            \noalign{\smallskip}
            \hline
         \end{tabular}
\end{table}
\begin{table}[]{}
\leftline{(Table\,\ref{sptable} continued)}
\begin{tabular}{l@{ }l@{  }l@{  }l@{  }l@{  }}
\hline
\noalign{\smallskip}
     (1)               &(2)        &(3)         &(4)        &(5)   \\
\noalign{\smallskip}
\hline
\noalign{\smallskip}
                              &72400605           &0.14     &790.7   &7.0    \\
                              &72400702           &0.071    &794.2   &2.5    \\
\noalign{\smallskip}
\multicolumn{5}{l}{\textbf{Group E}}                             \\
\noalign{\smallskip}
$\astrobj{19039+0809}$        &31600514$^e$       &0.035    &45.6     &1.68  \\
$\astrobj{23412-1533}$        &20000913$^e$       &0.071    &67       &2.74  \\
\noalign{\smallskip}
\multicolumn{5}{l}{\textbf{Group U1}}                    \\
\noalign{\smallskip}
$\astrobj{17103-3702}$        &28901940           &0.07     &864.7   &10.2   \\
                              &48202504           &0.035    &857.2   &5.0    \\
                              &48903704           &0.035    &846.5   &4.4    \\
                              &49601204           &0.035    &868.5   &5.4    \\
                              &50301504           &0.035    &859.7   &5.7    \\
                              &51001804           &0.035    &865.6   &8.6    \\
                              &67100301           &0.07     &883.5   &7.0    \\
                              &67801601           &0.035    &885.7   &5.8    \\
$\astrobj{17411-3154}$        &46901615           &0.07     &1382    &5.0    \\
$\astrobj{18196-1331}$        &34000203           &0.071    &4476    &13.1   \\
$\astrobj{19114+0002}$        &31900901           &0.072    &642.3   &4.0    \\
                              &52500806$^m$       &0.035    &660.4   &2.7    \\
                              &52500861           &0.29     &655.1   &2.9    \\
$\astrobj{22176+6303}$        &09101821$^*$       &0.07     &350.8   &8.8    \\
                              &60101805           &0.067    &11285   &28.0   \\
                              &82301120           &0.071    &11743   &85.8   \\
                              &82301122           &0.07     &11771   &99.7   \\
                              &82301123           &0.071    &11792   &39.5   \\
\noalign{\smallskip}
\multicolumn{5}{l}{\textbf{Group U2}}                            \\
\noalign{\smallskip}
$\astrobj{10197-5750}$        &10300135           &0.28     &634.7   &5.1    \\
$\astrobj{17439-2845}$        &69601311           &0.27     &1854    &15.6   \\
            \noalign{\smallskip}
            \hline
         \end{tabular}
\end{table}
\begin{table}[]{}
\leftline{(Table\,\ref{sptable} continued)}
\begin{tabular}{l@{ }l@{  }l@{  }l@{  }l@{  }}
\hline
\noalign{\smallskip}
     (1)               &(2)        &(3)         &(4)   &(5)        \\
\noalign{\smallskip}
\hline
\noalign{\smallskip}
\multicolumn{5}{l}{\textbf{Group U3}}                            \\
\noalign{\smallskip}
$\astrobj{01037+1219}$        &57700513           &0.07     &163.7   &3.5    \\
                              &57701103           &0.07     &170.4   &2.4    \\
$\astrobj{01304+6211}$        &61300914           &0.07     &141.8   &3.4    \\
$\astrobj{05506+2414}$        &83901512           &0.07     &112.3   &2.5    \\
$\astrobj{15452-5459}$        &48800916           &0.07     &339.5   &9.3    \\
$\astrobj{17150-3224}$        &32702239           &0.14     &294     &5.1    \\
$\astrobj{18050-2213}$        &33100802           &0.071    &290.5   &4.4    \\
$\astrobj{18198-1249}$        &47801312           &0.14     &63.9    &1.4    \\
$\astrobj{18498-0017}$        &32300501           &0.071    &173.7   &4.6    \\
$\astrobj{18560+0638}$        &70900321           &0.071    &128.7   &3.3    \\
$\astrobj{19343+2926}$        &35501620$^m$       &0.14     &86.8    &2.9    \\
                              &36701902           &0.071    &122.2   &4.6    \\
                              &52000720$^m$       &0.14     &131.4   &3.7    \\
                              &52000845           &0.071    &126.6   &2.6    \\
$\astrobj{22177+5936}$        &28300920           &0.071    &94.2    &3.0    \\
\noalign{\smallskip}
\multicolumn{5}{l}{\textbf{Group U4}}                            \\
\noalign{\smallskip}
$\astrobj{07027-7934}$        &14101004           &0.14     &45.8    &2.5    \\
                              &56700908$^m$       &0.035    &29.2    &2.3    \\
                              &56700981           &0.28     &43.3    &2.1    \\
                              &72200934           &0.072    &24.4    &17.2   \\
$\astrobj{10580-1803}$        &22800816$^m$       &0.071    &78.7    &4.4    \\
$\astrobj{16280-4008}$        &08402635           &0.071    &107.2   &4.6    \\
$\astrobj{18272+0114}$        &14900719$^m$       &0.035    &93.7    &3.1    \\
$\astrobj{18437-0302}$        &71501714           &0.071    &96.9    &4.1    \\
$\astrobj{18596+0315}$        &49901207           &0.14     &32.5    &1.4    \\
$\astrobj{19219+0947}$        &54700310           &0.07     &44.7    &2.9    \\
$\astrobj{19255+2123}$        &17600528           &0.071    &48.6    &2.4    \\
            \noalign{\smallskip}
            \hline
         \end{tabular}
\end{table}
\begin{table}[]{}
\leftline{(Table\,\ref{sptable} continued)}
\begin{tabular}{l@{ }l@{  }l@{  }l@{  }l@{  }}
\hline
\noalign{\smallskip}
     (1)               &(2)        &(3)         &(4)   &(5)        \\
\noalign{\smallskip}
\hline
\noalign{\smallskip}
$\astrobj{20077-0625}$        &34400719           &0.072    &207.8   &9.1    \\
$\astrobj{22036+5306}$        &54800798           &0.07     &119     &6.2    \\
\noalign{\smallskip}
\multicolumn{5}{l}{\textbf{Group U5}}                            \\
\noalign{\smallskip}
$\astrobj{17463-3700}$        &32400610           &0.27     &6.4     &4.0    \\
$\astrobj{20000+4954}$        &26300417$^m$       &0.14     &2.2     &1.8    \\
            \noalign{\smallskip}
            \hline
         \end{tabular}
\begin{list}{}{}
\scriptsize
\item[Notes for column 2:]
\item[`*': ] The down scan of 09901026 is of bad quality, only its up scan is used. 
There are data from different rasters for 09101821, 63300601, 63300602, the data of 
raster 3, 1, 1 are used respectively.\\
\item[`m': ]Several spectra are marked by `m': 52500806, 47201318, 35501620, 52000720, 
56700908, 22800816, 14900719 and 26300417 were observed in LWS 02 mode; 
55000813, 55500942, 36401611  were observed in LWS 04 mode. The other 
spectra were observed in LWS 01 mode.\\
\item[`a, t, e': ]Spectra marked by `a' means the 53.3\,$\mu$m line is detected, 
by `t' means only tentatively detected and `e' means dominated by spurious 
broad emission feature around 53.3\,$\mu$m.
\end{list}
\end{table}

The ISO Spectral Analysis Package (ISAP2.1)\footnote{The ISO Spectral Analysis Package (ISAP) is a joint development by the LWS and SWS Instrument Teams and Data Centers. Contributing institutes are CESR, IAS, IPAC, MPE, RAL and SRON.}, together with LWS Interactive Analysis software (LIA10.0), is used to process and analyse all these spectra. Glitches remained in the Auto Analysis Results (AARs) are removed by hand carefully. Small memory effect is neglected by directly averaging together the two scans. When large memory effect is present or prominent discrepancy exits between the up and down scans, the two scans are averaged separately to confirm the detection of the 53.3\,$\mu$m line feature.

Here is a brief description of the ISO spectra used in this paper. Some of the OH/IR sources (12 sources out of 47) have more than one ISO LWS associations. For most of these OH/IR sources, all their associated ISO LWS spectra have been processed and considered; the only three exceptions are \astrobj{IRAS\,17424-2859}, \astrobj{IRAS\,17441-2822} and \astrobj{IRAS\,20255+2712}, which have lots of repeatedly observed LWS spectra and only part of these ISO LWS spectra are processed and presented in Table\,\ref{sptable}. Therefore, only 102 ISO LWS spectra associated with the 47 Galactic OH/IR sources are listed in Table\,\ref{sptable}. 

For the purpose of comparison, the ISO LWS spectra of two OH megamasers, \astrobj{Arp\,220} and \astrobj{NGC\,253} are also processed. The ISO LWS or SWS spectra  have been discussed by Skinner et al.\,(\cite{ski97}) and Bradford et al.\,(\cite{bra99}) respectively. Totally 3 ISO LWS spectra for \astrobj{Arp\,220} and 1 spectrum for \astrobj{NGC\,253} are found covering the 53.3\,$\mu$m region (see Table\,\ref{sptable}). Hence, there are totally 49 OH maser sources with 106 associated ISO spectra listed in Table\,\ref{sptable}. 

ISO LWS observation could be performed in one of the four different possible modes: LWS 01, 02, 03 or 04, corresponding to raster scanning of the full working wavelength range ($43\sim196.7\,\mu$m) or individual small line ranges, Fabry-P${\acute e}$rot (FP) high resolution scanning of specified wavelength ranges or individual small line ranges respectively. Among the ISO spectra used in this paper, there are only eight LWS 02 and three LWS 04 spectra while all others are observed in LWS 01 mode. For each observing mode, the actual spectral resolution depends on the scan speed. For the 106 LWS spectra used in this paper, the spectral resolution around 53.3\,$\mu$m is found to be about 0.0015 (2 spectra) or 0.068\,$\mu$m (1 spectrum) for LWS 04 mode, 0.035 (3 spectra), 0.07 (2 spectra) or 0.14\,$\mu$m (3 spectra) for LWS 02 mode, 0.035 (10 spectra), 0.07 (66 spectra), 0.14 (12 spectra) or 0.28\,$\mu$m (7 spectra) for LWS 01 mode. Details of the ISO LWS instrument and observations can be found in Clegg et al.\,(\cite{cle96}).

\section{Detectability of the 53.3\,$\mu$m absorption line}
\label{detectability}

\subsection{Statistics upon the spectra}
\label{statonsp}

Those spectra with the 53.3\,$\mu$m line feature detected and tentatively detected can be defined as group `a' and `t' and are marked in Table\,\ref{sptable} by superscripts `a' and `t' respectively; the other non--detections can be defined as group `u'. Another two spectra with spurious broad emission feature around 53.3\,$\mu$m are defined as group `e' and marked by superscript `e' in Table\,\ref{sptable}. The tentative detections are defined as those spectra with measured line depth/height smaller than 3 times of the sigma noise level of the continuum flux. The spurious emission-like feature in group-e spectra is, according to the ISO LWS handbook, caused by a near-infrared light leakage in the blocking filters located in front of the detectors. Among the 106 ISO LWS spectra, there are 21 group-a, 24 group-t, 2 group-e and 59 group-u ones. The group-u can be further divided according to their 53.3\,$\mu$m continuum flux 3 sigma noise level into 5 sub-groups: u1, u2, u3, u4 and u5 with the 3 sigma noise level $<\,2\%$, $<\,4\%$, $<\,10\%$, $<\,24\%$ and $>\,24\%$ of the continuum flux level respectively. Totally there are 15 group-u1, 9 group-u2, 16 group-u3, 14 group-u4 and 5 group-u5 spectra. The group-u1 and u2 spectra are considered as highly probable non--detection cases for the 53.3\,$\mu$m line while the other group-u3, u4 and u5 spectra can be considered as noisy spectra. Therefore, the detection and non--detection rates based on the counting of spectra can be estimated to be $43\%$ and $23\%$ respectively, if the group-t spectra are considered as detections and the group-e spectra are excluded. 

Here we do not defined non--detections as we did in paper I where we compared the 3 sigma noise level with the expected 34.6\,$\mu$m line depth estimated from the blue peak intensity of the OH 1612\,MHz maser because: (1) no theoretical hint about the quantitative relationship between the 53.3\,$\mu$m line and the OH 1612\,MHz maser line strength is available; (2) the filling emission feature widely appearing in the 53.3\,$\mu$m line profile, as we will see in the following discussions, makes it meaningless to compare the directly measured 53.3\,$\mu$m line depth/height with the OH maser line strength. 

\begin{figure}[]
   \centering
	 \includegraphics[width=9cm]{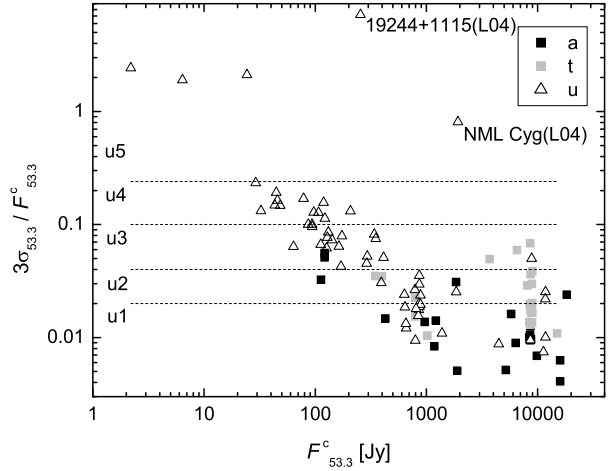}
   \caption{The relative 53.3\,$\mu$m continuum 3 sigma noise level against the continuum flux. The horizontal lines delineate how the group-u spectra are further grouped into u1, u2, u3, u4 and u5.}
   \label{sig-cont}
   \end{figure}

The above classification is visualized in Fig.\,\ref{sig-cont} in which the horizontal lines delineate how the group-u is further divided into subgroups. This figure also shows that the detection of the 53.3\,$\mu$m line feature generally occurs in spectra with low 3 sigma noise and high continuum flux, which indicates that the selection effect produced by the limited sensitivity of ISO LWS spectrometer ({\em sensitivity selection effect}) is serious in our sample.  Therefore, some weak non--detections could be explained by the insufficient sensitivity of the instrument. Note that two non--detection spectra marked by their source names in the figure show abnormally high noise; they are both LWS 04 spectra. Maybe the ISO LWS 04 mode is sometimes not as reliable as the other modes.

\begin{figure}[]
   \centering
	 \includegraphics[width=9cm]{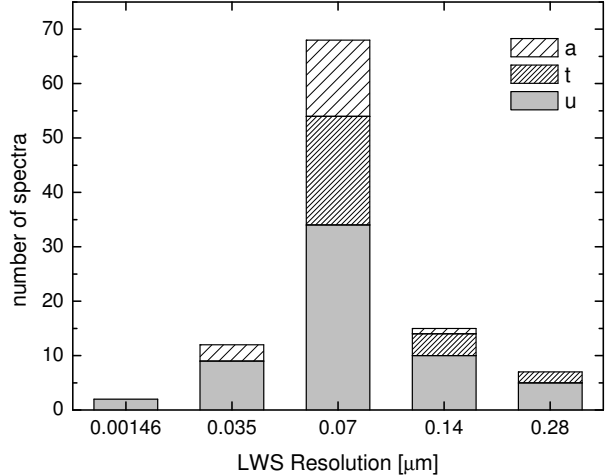}
   \caption{Stack bar figure of spectra of different groups with different resolutions.}
   \label{stResolv}
   \end{figure}

Our sample of ISO spectra is biased by selection effects because the observations were originally designed for quite different purposes by different observers. Beside the {\em sensitivity selection effect} mentioned above, another prominent selection effect in our sample spectra is the different resolutions used for different observations. As Fig.\,\ref{stResolv} shows, most of the spectra were observed with a resolution of 0.07\,$\mu$m while a non--negligible number of observations are performed with higher or lower resolutions ({\em resolution selection effect}). The different resolutions will affect the strength of spectral features and hence affect the detectability of the 53.3\,$\mu$m line. 

The detectability of the 53.3\,$\mu$m OH line feature depends on several facts such as noise level, resolution and the genuine strength of the line feature itself. As described above, our sample of spectra are affected by instrumental sensitivity and resolution selection effects. If the spectra either with 3 sigma noise level higher than $4\%$ or with resolution other than 0.07\,$\mu$m are excluded, we can define a relatively more homogeneous sample\,II with 43 spectra in it. Based on  sample\,II another detection and non--detection rate can be re-estimated to be $74\%$ (32 spectra out of 43) and $26\%$ (11 spectra out of 43) respectively. We do not further constrain the detection and non--detection rate upon the effect of the genuine 53.3\,$\mu$m OH line strength because it is difficult to estimate an expected line strength for this OH absorption line due to the reasons described earlier.

\subsection{Statistics upon OH/IR sources}
\label{statonobj}

The OH/IR sources associated with these ISO LWS spectra can be grouped in a similar way. Those OH/IR sources with at least one group-a spectrum are defined as group `A'; those with no group-a but at least one group-t spectrum are defined as group `T'; those with neither group-a nor group-t spectra are defined as group `U' and they can be further grouped into U1, U2, U3, U4 and U5 if their associated best quality spectrum belongs to group u1, u2, u3, u4 and u5 respectively. Another two OH/IR sources with their only one spectrum showing spurious broad emission around 53.3\,$\mu$m are classified as group `E'. As a result, the 47 galactic OH/IR sources are divided into 10 group-A, 5 group-T, 5 group-U1, 2 group-U2, 11 group-U3, 10 group-U4, 2 group-U5 and 2 group-E sources; the two megamasers, \astrobj{Arp\,220} and \astrobj{NGC\,253}, both belong to group `A'. 

Another detection and non--detection rate of the 53.3\,$\mu$m line can be estimated, by counting group-A, T, U galactic OH/IR sources, to be 33$\%$  (15 sources out of 45) and 16$\%$(7 sources out of 45) respectively. Here again, group-T sources are considered as detections and group-E sources are excluded. However these rates are also affected by selection effects, as decribed below.
\begin{figure}[]
   \centering
	 \includegraphics[width=9cm]{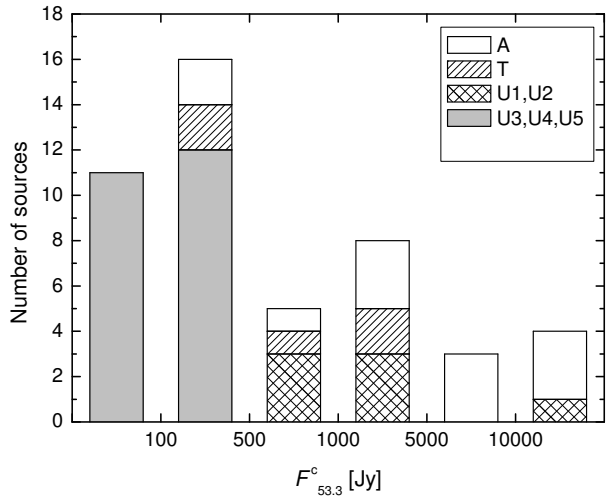}
   \caption{Stack bar figure of  OH/IR sources of different groups with different 53.3\,$\mu$m continuum flux.}
   \label{statcont53}
   \end{figure}

Beside the instrument related selection effects discussed in Sect.	\,\ref{statonsp} that have significantly reduced the number of useful spectra in our sample, the distribution of our sample sources upon their brightness and IR color can be another two potential selection effects, just as we discussed in Paper\,I. The statistics of all our sample sources upon 53.3\,$\mu$m continuum flux are plotted in Fig.\,\ref{statcont53}. The source counting distribution in this figure seems to decrease from low flux to high flux slightly irregularly, which implies our sample is incomplete ({\em brightness selection effect}). Another indication from the figure is that most group-A, T, U1 and U2 sources are located in high flux groups only and group-U3, U4 and U5 sources (noisy sources) only concentrate in the two lowest flux groups, which implies that the {\em brightness selection effect}, together with the instrument related {\em sensitivity selection effect}, has largely affected the detection/non--detection rate estimated above. The other potential selection effect, {\em color selection effect}, can be checked for in the IRAS two color diagram (wherein the IRAS color indices are defined as $C_{21}$=log$(F_{25}/F_{12})$, $C_{32}$=log$(F_{60}/F_{25})$) in Fig.\,\ref{IRAScc} in which our sample sources distribute in almost all areas where OH masers may appear. Therefore the {\em color selection effect} is not a serious problem in our sample.

\section{ISO Infrared color}
\label{ISOIRc}

An ISO infrared color $C_{ISO}$ can be defined using the 53.3\,$\mu$m continuum flux $F^{c}_{53.3}$ and the 34.6\,$\mu$m continuum flux $F^{c}_{34.6}$ from Paper\,I, i.e., $C_{ISO}$ = log($F^{c}_{53.3}/F^{c}_{34.6}$). For those OH/IR sources with multiple ISO LWS observations, the $F^{c}_{53.3}$ of most of the reliable spectra listed in Table\,\ref{sptable} are used to estimate a mean value. Totally three spectra: TDT 36401611 of \astrobj{IRAS 19244+1115}, TDT 35501620 of \astrobj{IRAS 19343+2926} and TDT 09101821 of \astrobj{IRAS 22176+6303} are rejected when calculating the mean continuum flux because their continuum fluxes are obviously smaller than that of the other spectra associated with the same OH/IR sources. The reason for the too low fluxes may be related to the ill performance of the ISO instruments. The resulted mean continuum fluxes are listed in column (4) of Table\,\ref{objtable}. Also listed in this table are columns: (1) -- source name; (2) -- SIMBAD object class; (3) IRAS/LRS classification defined by Volk and Cohen\,(\cite{vol89}) and given by Kwok et al.\,(\cite{kwo97}); (5) -- blue peak flux of 1612\,MHz OH maser ($F^{p,blue}_{OH}$) and (6) -- reference code for OH peak flux and some notes. 
%
\begin{table}[]{}
\caption[]{ISO 53.3\,$\mu$m and OH 1612\,MHz maser spectral quantities of OH/IR sources. 
These objects are grouped in the same way as Table\,\ref{sptable}. {\it F}$^{c}_{53.3}$ 
in this table is a mean value for each OH/IR source.}
\label{objtable}
\begin{tabular}{l@{ }l@{  }l@{  }l@{  }l@{  }l@{  }}
\hline
\noalign{\smallskip}
      Name             &Class      &LRS         &$F^{c}_{53.3}$[Jy] &$F^{p,blue}_{OH}$[Jy]  &ref\\
     (1)               &(2)        &(3)         &(4)                    &(5)                 &(6)\\
\noalign{\smallskip}
\hline
\noalign{\smallskip}
\multicolumn{6}{l}{\textbf{Group A}}                                 \\
\noalign{\smallskip}
$\astrobj{03507+1115}$        &Mira     &E        &427     &2.2     &tLVHW  \\
$\astrobj{06053-0622}$        &Comp     &H        &15912   &2.03    &tL-HPhD\\
$\astrobj{07209-2540}$        &PV*      &E        &1890    &200     &tLVHW  \\
$\astrobj{17424-2852}$        &V*       &I        &5182    &0.12    &tLVHW  \\
$\astrobj{17424-2859}$        &IR       &H        &18284   &0.36    &tLVHW  \\
$\astrobj{17430-2848}$        &Cl*      &H        &1847    &0.95    &LWH92  \\
$\astrobj{17441-2822}$        &IR       &-        &6211    &1.39    &ATCAb  \\
$\astrobj{17574-2403}$        &H\,II    &H     	&15442   &34.7    &tLVHW  \\
$\astrobj{20255+3712}$        &H\,II    &H     	&8716    &0.22    &AOe(fig)    \\
$\astrobj{Arp\,220}$          &IntG     &-        &118.4   &31.4    &Baa87(fig)(1p)     \\
$\astrobj{NGC\,253}$          &SeyfG    &P        &964.7   &----    &       \\
$\astrobj{NML\,Cyg}$          &V*       &-        &1438    &638     &(priv) \\
\noalign{\smallskip}
\multicolumn{6}{l}{\textbf{Group T}}                                 \\
\noalign{\smallskip}
$\astrobj{06319+0415}$        &IR       &H        &1019    &0.26    &tLVHW(1p)\\
$\astrobj{16342-3814}$        &pAGB     &H        &346.5   &7.1     &SAP93 \\
$\astrobj{17431-2846}$        &IR       &H        &3697    &0.21    &tLVHW \\
$\astrobj{18348-0526}$        &Mira     &A        &401.3   &195     &tL-H  \\
$\astrobj{19244+1115}$        &pAGB     &E        &796.4$^{*}$&45.6 &tLVHW \\
\noalign{\smallskip}
\multicolumn{6}{l}{\textbf{Group E}}                             \\
\noalign{\smallskip}
$\astrobj{19039+0809}$        &Mira     &E        &45.6    &4       &tLVHW\\
$\astrobj{23412-1533}$        &Mira     &E        &67      &0.06    &ISH94(1p)\\
\noalign{\smallskip}
\multicolumn{6}{l}{\textbf{Group U1}}                    \\
\noalign{\smallskip}
$\astrobj{17103-3702}$        &PN       &L        &866.4   &0.2     &ZLP89 \\
$\astrobj{17411-3154}$        &OH/IR    &A        &1382    &224     &tL-H      \\
$\astrobj{18196-1331}$        &YSO      &A        &4476    &2       &tL-HPhD  \\
            \noalign{\smallskip}
            \hline
         \end{tabular}
\end{table}
\begin{table}[]{}
\leftline{(Table\,\ref{objtable} continued)}
\begin{tabular}{l@{ }l@{  }l@{  }l@{  }l@{  }l@{  }}
\hline
\noalign{\smallskip}
     (1)               &(2)        &(3)         &(4)        &(5)        &(6)\\
\noalign{\smallskip}
\hline
\noalign{\smallskip}
$\astrobj{19114+0002}$        &PAGB     &H        &652.6   &3.98    &L89   \\
$\astrobj{22176+6303}$        &H\,II    &H     	&11648$^{*}$&0.34 &BLS90(fig)(1p)\\
\noalign{\smallskip}
\multicolumn{6}{l}{\textbf{Group U2}}                            \\
\noalign{\smallskip}
$\astrobj{10197-5750}$        &WR*      &H        &634.7   &44       &tLVHW \\
$\astrobj{17439-2845}$        &OH/IR    &H        &1854    &4.1      &tL-H  \\
\noalign{\smallskip}
\multicolumn{6}{l}{\textbf{Group U3}}                            \\
\noalign{\smallskip}
$\astrobj{01037+1219}$        &OH/IR    &E        &167.0   &48.2     &tLVHW\\
$\astrobj{01304+6211}$        &Mira     &A        &141.8   &32       &tLVHW    \\
$\astrobj{05506+2414}$        &HH       &H        &112.3   &0.18     &AOa(fig)\\
$\astrobj{15452-5459}$        &pAGB     &U        &339.5   &12.3     &tL-H     \\
$\astrobj{17150-3224}$        &pAGB     &H        &294     &2.98     &tL-H(1p) \\
$\astrobj{18050-2213}$        &SRPV*    &E        &290.5   &16       &tLVHW    \\
$\astrobj{18198-1249}$        &MlCl     &A        &63.9    &7.4      &tLVHW\\
$\astrobj{18498-0017}$        &Mira     &H        &173.7   &29.5     &tLVHW    \\
$\astrobj{18560+0638}$        &Mira     &A        &128.7   &20       &tLVHW    \\
$\astrobj{19343+2926}$        &Em*      &H        &126.7$^{*}$ &0.45 &tLVHW    \\
$\astrobj{22177+5936}$        &V*       &A        &94.2    &41       &tLVHW    \\
\noalign{\smallskip}
\multicolumn{6}{l}{\textbf{Group U4}}                            \\
\noalign{\smallskip}
$\astrobj{07027-7934}$        &PN       &H        &35.7    &8.62     &tL-H      \\
$\astrobj{10580-1803}$        &SRPV*    &E        &78.7    &2        &EN79      \\
$\astrobj{16280-4008}$        &PN       &L        &107.2   &0.13     &LC96(fig)\\
$\astrobj{18272+0114}$        &V*O      &-        &93.7    &0.069    &AOe(fig) \\
$\astrobj{18437-0302}$        &IR       &-        &96.9    &0.80     &BLW94(1p) \\
$\astrobj{18596+0315}$        &OH/IR    &-        &32.5    &6.64     &ATCA3     \\
$\astrobj{19219+0947}$        &PN       &H        &44.7    &5.8      &tLVHW     \\
$\astrobj{19255+2123}$        &PN       &H        &48.6    &0.4      &tLVHW     \\
$\astrobj{20077-0625}$        &V*       &E        &207.8   &11.0     &tL-H      \\
$\astrobj{22036+5306}$        &OH/IR    &U        &119     &0.7      &tLH91     \\
            \noalign{\smallskip}
            \hline
         \end{tabular}
\end{table}
\begin{table}[]{}
\leftline{(Table\,\ref{objtable} continued)}
\begin{tabular}{l@{ }l@{  }l@{  }l@{  }l@{  }l@{  }}
\hline
\noalign{\smallskip}
     (1)               &(2)        &(3)         &(4)   &(5)        &(6)\\
\noalign{\smallskip}
\hline
\noalign{\smallskip}
\multicolumn{6}{l}{\textbf{Group U5}}                            \\
\noalign{\smallskip}
$\astrobj{17463-3700}$        &PN       &F        &6.4     &0.24     &ISH94(1p)\\
$\astrobj{20000+4954}$        &Mira     &E        &2.2     &3.8      &tLVHW    \\
            \noalign{\smallskip}
            \hline
         \end{tabular}
\begin{list}{}{}
\scriptsize
\item[Column 2:]

Cl* = `Cluster of Stars',                     
Comp = `Composite object',                    
Em* = `Emission Line Star',                   
HH = `Herbig-Haro object',                    
IntG = `Interacting Galaxies',                
IR = `Infra-Red source',                      
Mira = `Variable star of Mira Cet type',      
OH/IR = `Star with envelope of OH/IR type',   
pAGB = `Post-AGB Star',                       
PN = `Planetary Nebula',                      
PV* = `Pulsating variable Star',         
SeyfG = `Seyfert Galaxy',                
SRPV* = `Semi-regular Pulsating Star',   
V* = `Variable Star',                    
V*O = `Variable Star of Orion type',     
WR* = `Wolf-Rayet Star',                 
YSO = `Young Stellar Object',            
H\,II = `H\,II (ionized) region',        
MlCl = `Molecular Cloud'.       
\item[Column 4:]`*': Following spectra have been excluded due to their abnormal flux value when 
calculating mean continuum flux: TDT36401611 (\astrobj{19244+1115}), TDT35501620 (\astrobj{19343+2926}), 
TDT09101821 (\astrobj{22176+6303}).
\item[Column 6:]Reference codes are defined as follow while `(fig)' means the peak flux is measured 
from the published maser profile figure and `(1p)' means it is a single peak maser.\\
         \begin{tabular}{l@{ }l@{  }}
                AOe: Lewis(\,\cite{lew94})                  &LC96: te Lintel Hekkert \& Chapman(\,\cite{tel96})\\      
                ATCA3: Sevenster et al.(\,\cite{sev01})     &LWH92: Lindqvist et al.(\,\cite{lin92})           \\      
                ATCAb: Sevenster et al.(\,\cite{sev97b})    &SAP93: Silva et al.(\,\cite{sil93})               \\      
                BLS90: Braz et al.(\,\cite{bra90})          &ZLP89: Zijlstra et al.(\,\cite{zij89})            \\      
                BLW94: Blommaert et al.(\,\cite{blo94})     &tL--H: te Lintel Hekkert et al.(\,\cite{tel91a})    \\     
                Baa87: Baan \& Hachick(\,\cite{baa87})      &tL--HPhD: te Lintel Hekkert's PhD thesis (\,\cite{tel90})\\
                EN79: Engels(\,\cite{eng79})                &tLH91: te Lintel Hekkert(\,\cite{tel91b})       \\        
                ISH94: Ivison et al.(\,\cite{ivi94})        &tLVHW: te Lintel Hekkert et al.(\,\cite{tel89})\\         
                L89: Likkel(\,\cite{lik89})                 &(priv): private communication.       \\                   
         \end{tabular}
\end{list}
\end{table}

\begin{figure}[]
   \centering
	 \includegraphics[width=9cm]{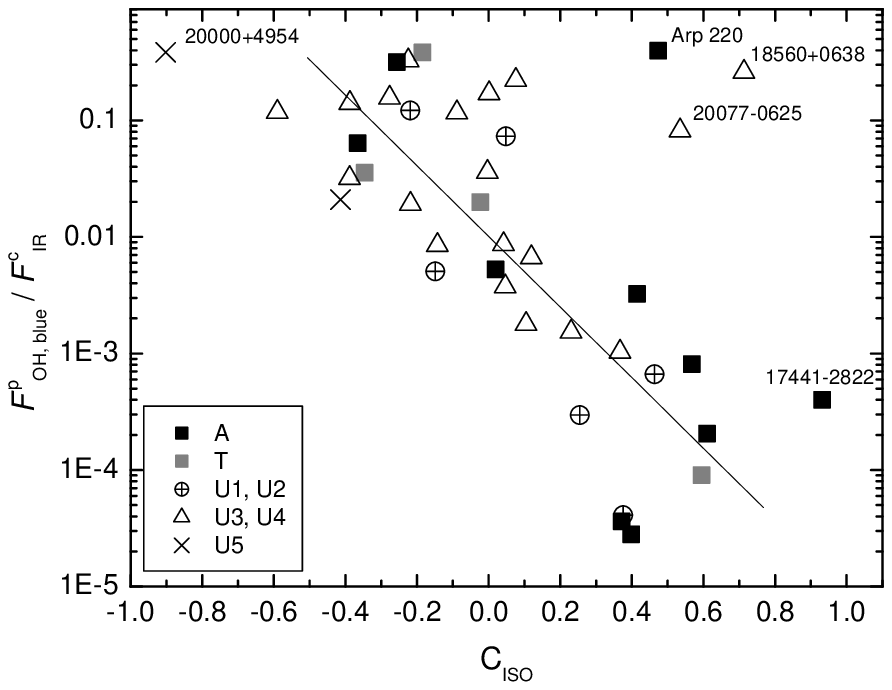}
   \caption{Blue peak flux of 1612\,MHz OH maser against ISO IR color. Outliers are labelled by their source names.}
   \label{ISOcSOH}
   \end{figure}

The relation between the blue peak flux of 1612\,MHz OH maser ($F^{p,blue}_{OH}$) and $C_{ISO}$ is plotted in Fig.\,\ref{ISOcSOH} in which $F^{p,blue}_{OH}$ is divided by a mean ISO IR flux $F^{c}_{IR}$ defined as the mathematic mean of $F^{c}_{34.6}$ and $F^{c}_{53.3}$ to remove the effect of different distance to the sources. (When $F^{c}_{34.6}$ or $F^{c}_{53.3}$ is unavailable, the other quantity is used for $F^{c}_{IR}$.)  This maser to IR flux ratio can be considered as some kind of measure of OH maser `efficiency'. Most of our sample OH/IR sources, irrespect to whether the 53.3\,$\mu$m line is detected or not, distribute along a uniform inverse correlation between maser efficiency and $C_{ISO}$ except five outliers labelled by their source names in the figure. A line is fitted to the data not labelled in fig.\,\ref{ISOcSOH}.
   \begin{figure}[h]
   \centering
	 \includegraphics[width=9cm]{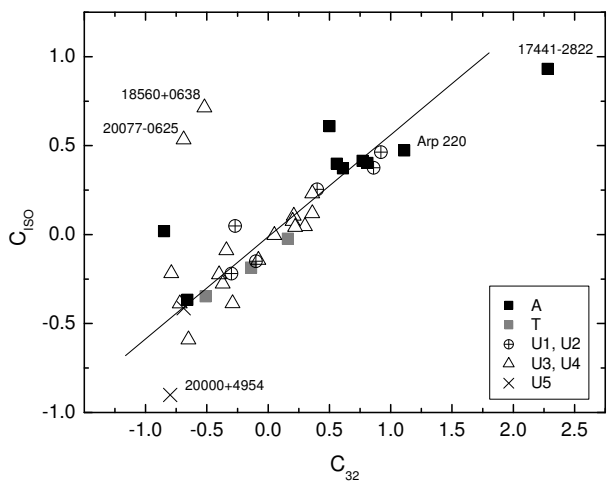}
   \caption{ISO IR color against IRAS color. Sources labelled in Fig.\,\ref{ISOcSOH} are also labelled in this figure by their source names.}
   \label{ISOcIRASc}
   \end{figure}

Among the five outliers in Fig.\,\ref{ISOcSOH}, \astrobj{Arp\,220} is a megamaser and its too red ISO IR color may be due to IR emission from large volumns of cold interstellar dust. The too blue color of \astrobj{IRAS 20000+4954} and too red color of the other three sources may be due to errors in their ISO continuum flux. This point can be checked for by comparing $C_{ISO}$ with IRAS color $C_{32}$ because the wavelength of $F^{c}_{34.6}$ is close to that of $F_{25}$ and the wavelength of $F^{c}_{53.3}$ is close to that of $F_{60}$. The correlation of the two colors is plotted in Fig.\,\ref{ISOcIRASc} for sources with good IRAS flux data (with quality factor Q$>$1). The five sources labelled in Fig.\,\ref{ISOcSOH} are also labelled in the figure. As seen from the figure, $C_{ISO}$ of most of our sample sources obeys a uniform linear correlation with $C23$ except four of the labelled sources (outliers). A solid line is fitted to those good points (unlabelled ones and \astrobj{Arp 220}) in the figure. A careful check of ISO positions against IRAS and SIMBAD positions shows that, for \astrobj{IRAS 18560+0638} and \astrobj{IRAS 20077-0625}, the too red ISO IR color is due to a positional error of the 34.6\,$\mu$m ISO observations which results in a too weak 34.6\,$\mu$m continuum flux. For \astrobj{IRAS 17441-2822} (=\astrobj{Sgr B2}), its positions in Fig.\,\ref{ISOcSOH} and \ref{ISOcIRASc} seem both not very far from the correlations, however we note that the ISO color in Fig.\,\ref{ISOcSOH} is larger than the correlation required while the IRAS color is even larger than the ISO color in fig.\,\ref{ISOcIRASc}, hence this source, if plotted in pump efficiency--IRAS color relation (not shown), will drift further away from the expected correlation. Position check shows that the 34.6\,$\mu$m observation is pointed at a position $\sim 35^{\prime\prime}$ away from the IRAS position. On the other hand the ISO beam size of about 80$^{\prime\prime}$ is larger than its IRAS error ellipse of $23^{\prime\prime}\times 6^{\prime\prime}$. Therefore the difference in observational position and beam size may account for part of the difference between the ISO and IRAS colors. However, the reason for the deviation of this source in Fig.\,\ref{ISOcSOH} is still not clear after our simple consideration because this source is located near the complicated crowded galactic center. (Interested readers are referred to Goicoechea and Cernicharo\,(\cite{goi02}) for recent works on the modelling of plentiful ISO spectra of this source) For the other source under the solid line, \astrobj{IRAS 20000+4954}, the ISO position agrees quite well with its IRAS and SIMBAD positions, but the ISO spectrum around 53.3\,$\mu$m is quite weak and noisy and hence the ISO 53.3\,$\mu$m flux is problematic. In a summary, the abnormally too red or too blue ISO IR color of three of the outliers in Fig.\,\ref{ISOcSOH} can be explained by bad flux data while for \astrobj{IRAS 17441-2822}, a GC source, the situation is still unclear.

\begin{figure}[h]
   \centering
	 \includegraphics[width=9cm]{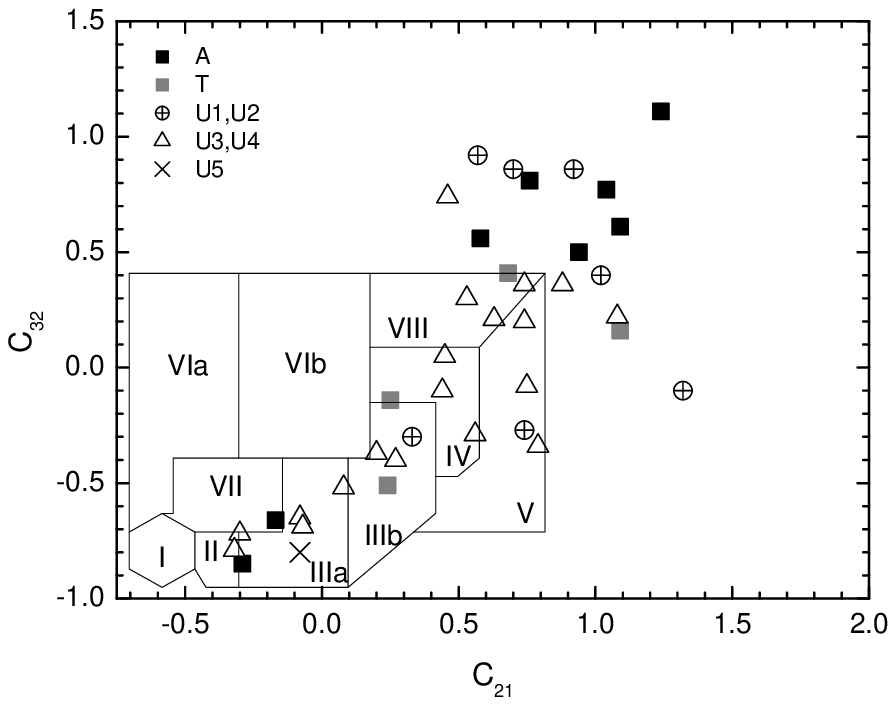}
   \caption{IRAS two color diagram for our sample of OH/IR sources. Only good quality IRAS fluxes (with Q$>$1) are used to make color indices. Regions are from van der Veen \& Habing~(\protect\cite{van88}).}
   \label{IRAScc}
	 \includegraphics[width=9cm]{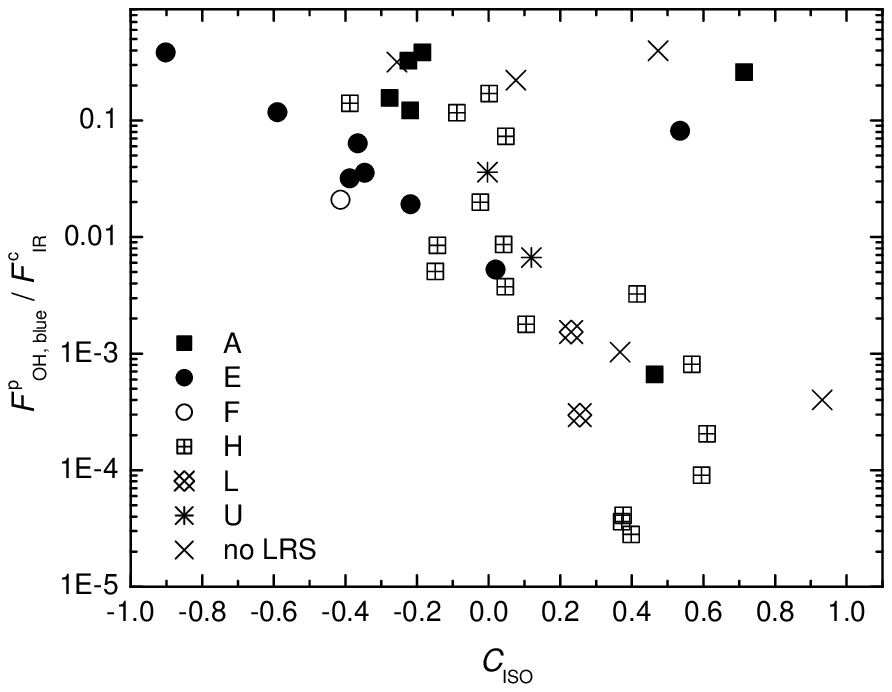}
   \caption{The same as Fig.\,\ref{ISOcSOH} but sources are differentiated by their IRAS/LRS types.}
   \label{ISOcSOHLRS}
   \end{figure}

The inverse correlation between $F^{p,blue}_{OH}$ and $C_{ISO}$ contradicts with that found by other authors such as Chen et al.~(\cite{che01}) or Sevenster~(\cite{sev02}). However the increase of maser efficiency with IRAS color found by these authors is built upon different sample of OH/IR sources. The correlation found by Chen et al.~(\cite{che01}) is based on OH/IR sources with IRAS/LRS type-A and E only. The correlation found by Sevenster~(\cite{sev02}) is based on a sample of double peak OH maser sources and hence OH masers in star formation regions are excluded from their sample. As seen from the IRAS two color diagram for our sample (see Fig.\,\ref{IRAScc}), it includes a large number of very red cool sources which may be H\,II regions. Hence the inverse correlation between $F^{p,blue}_{OH}$ and $C_{ISO}$ might be supported mainly by very cool sources such as H\,II regions. This idea is confirmed by Fig.\,\ref{ISOcSOHLRS} in which the sources are differentiated by their IRAS/LRS type, say, the inverse correlation is mainly supported by H type sources that are probably H\,II regions. The decrease of maser efficiency with IR color can be naturally explained by the increase of cool interstellar dust content around the cooler H\,II regions but outside of the maser region, for they increase the IR flux and IR color but contribute little to the pumping of the OH maser.

\section{The 53.3\,$\mu$m absorption profiles}
\label{profiles}

The detected 53.3\,$\mu$m features in group-a spectra are classified into six groups by comparing the spectral profiles by eye: (1) red-shifted absorption with blue-shifted filling emission ({\em RaBe}); (2) blue-shifted absorption with red-shifted filling emission ({\em BaRe}); (3) asymmetrical absorption profile of which the blue side of the line looks steeper than the red side and so the line center is shifted to shorter than 53.3\,$\mu$m ({\em Asymm}); (4) broad profile ({\em BroPf}); (5) spurious broad emission profile ({\em SpurEm}); (6) absorption line towards galactic center sources or megamaser galaxies, which is centered precisely at 53.3\,$\mu$m (after redshift correction for megamasers) and show left-right symmetrical broad profiles ({\em GCMega}). The spectral profile of NGC\,253 is peculiar, for it shows a very broad absorption tail at the blue side of the line, but it is included in the `{\em GCmega}' group for the convenience of description. 

\begin{figure}[]
   \centering
	 \includegraphics[width=9cm]{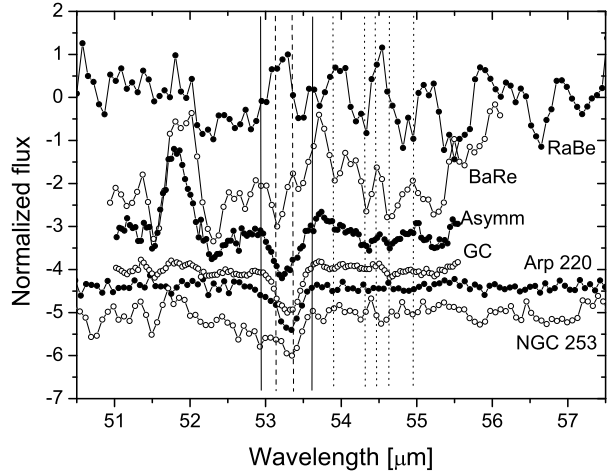}
   \caption{Representative normalized spectral profiles around 53.3\,$\mu$m. Some spectra are shifted downwards for clarity. The two solid vertical lines mark the wavelength range of the 53.3\,$\mu$m absorption line. The two dash lines mark the peak position of the blue filling emission of `{\em RaBe}' type spectrum and that of the red filling emission of the `{\em BaRe}' type spectrum. Another five dotted lines denote some emission-like or absorption-like features shared by different types of spectra. }
   \label{nnsp}
   \end{figure}

\begin{figure}[]
   \centering
\resizebox{\hsize}{!}{\includegraphics[angle=0]{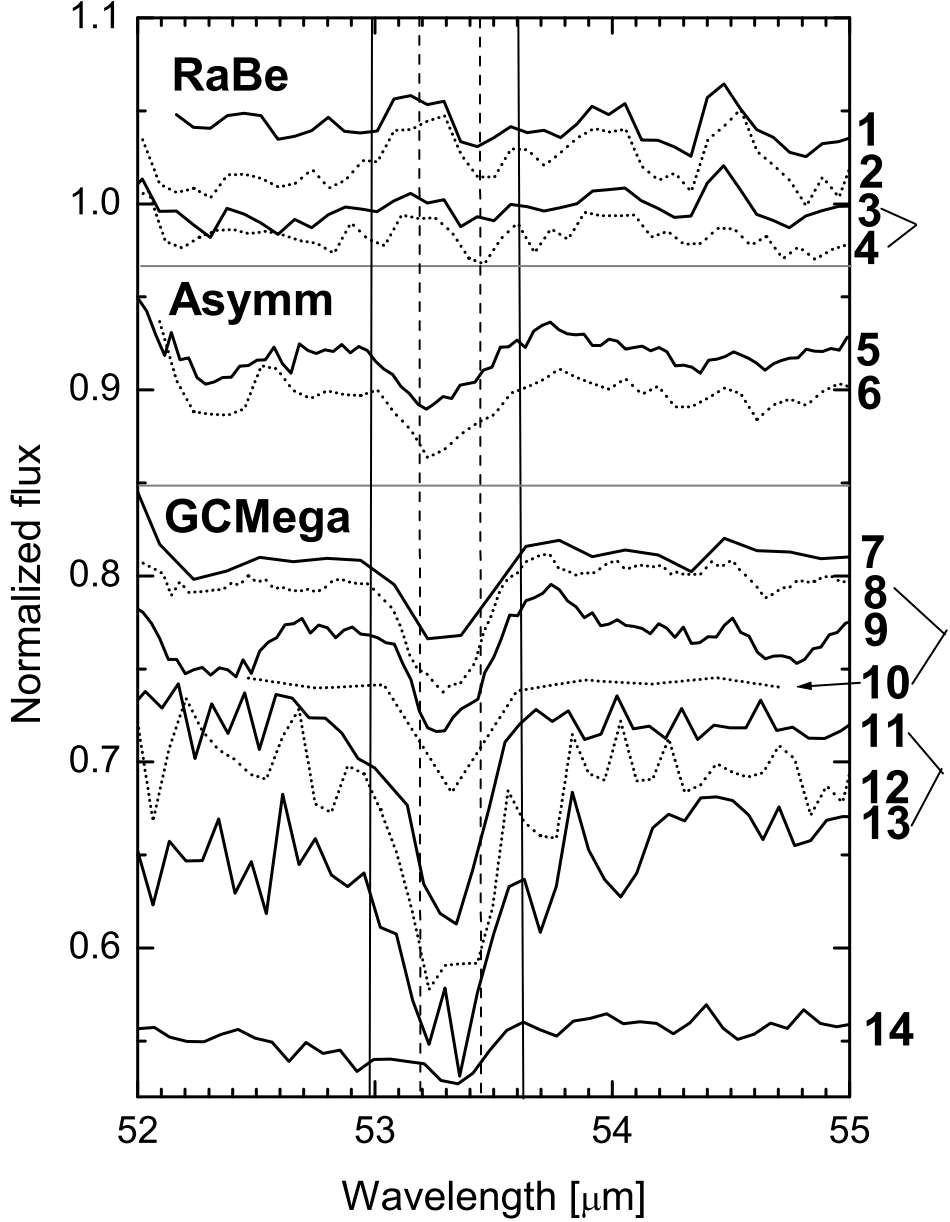}}
   \caption{Normalized spectral profiles around 53.3\,$\mu$m of `{\em RaBe}', `{\em Asymm}' and `{\em GCMega}' type spectra. The spectra are numerated at the right side of the figure and those spectra belonging to the same OH/IR source are grouped by two short lines. Two vertical full lines mark the edges of the 53.3\,$\mu$m absorption line and two vertical dashed lines mark the center of the filling emission and the residual absorption. Spectra are shifted upwards or downwards for clarity. From top to bottom they are: (1)65601707, (2)73502338, (3)34201304, (4)55500942, (5)32600904, (6)50701028, (7)49801004, (8)32601008, (9)32701306, (10)49400302, (11)27800202, (12)64000801, (13)64000916, (14)24701103}
   \label{groupsp1}
   \end{figure}

\begin{figure}[]
   \centering
	 \includegraphics[height=17cm, width=12cm]{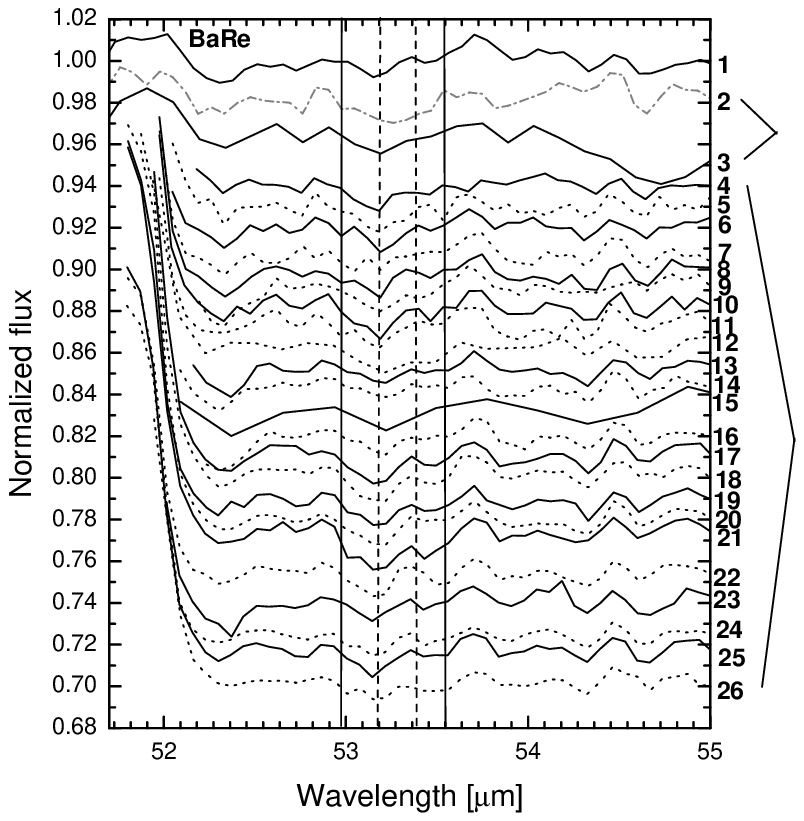}
   \caption{The same as Fig.\,\ref{groupsp1} but for `{\em BaRe}' type spectra. Most of these spectra belong to \astrobj{20255+3712}. Spectra No. (2) plotted in grey dash dot line belong to group `{\em BroPf}'. From top to bottom they are: (1)71002509, (2)09901026, (3)12500924, (4)01301603, (5)12600515, (6)13400331, (7)18204615, (8)20306815, (9)21002015, (10)22302315, (11)22302801, (12)35201327, (13)49602708, (14)52403008, (15)53000508, (16)53102508, (17)53803008, (18)54403605, (19)55205608, (20)57302606, (21)57902706, (22)58702901, (23)70601601, (24)76902601, (25)77602401, (26)86000201}
   \label{groupsp2}
   \end{figure}

In order to visualize the characteristics of the different profile groups, some representative spectra from each group except `{\em SpurEm}' and `{\em BroPf}', are shown in Fig.\,\ref{nnsp} while all group-a spectra are plotted in Fig.\,\ref{groupsp1} and Fig.\,\ref{groupsp2}. The two vertical full lines in Fig.\,\ref{nnsp} mark the range of the 53.3\,$\mu$m line, the two dashed lines inside them mark the center position of the blue filling emission of `{\em RaBe}' type spectra and that of the red filling emission of the `{\em BaRe}' type spectra while another set of dotted lines mark some nearby features shared among different groups. The vertical full lines and dashed lines in Fig.\,\ref{groupsp1} and Fig.\,\ref{groupsp2} are of similar use except that the two dashed lines are used to mark the center of the filling emission and the residual absorption. In Fig.\,\ref{groupsp2}, the only one `{\em BroPf}' type spectrum is included among `{\em BaRe}' type spectra (the light grey one (No. 2) in the figure) because its associated OH/IR source \astrobj{17574-2403} has another spectrum (No. 3) belonging to group `{\em BaRe}'. 

Fig.\,\ref{nnsp} demonstrates that, compared to the symmetrical profile of `{\em GCMega}' type spectra, the blue shifted filling emission in group `{\em RaBe}' and the red shifted filling emission in group `{\em BaRe}' look quite clear. The latter two groups also share emission-like features around 54.5\,$\mu$m, as shown by several dotted lines in the figure, but their profiles are not completely the same. Generally, the line features with filling emission component will become much weaker than the ones without filling emission component, and hence the detection of such weakened line feature may be quite uncertain due to the limited sensitivity and resolution of ISO LWS spectrometer. However, the special case of \astrobj{IRAS 20255+3712} gives us an important chance to confirm the detection of such a weak filling emission feature. This source has a lot of spectra (from No. 4 to 26 in Fig.\,\ref{groupsp2}) and the weak red-shifted filling emission feature is seen in almost all of these spectra. If the filling emission feature is unreal, it should not appear in so many repeated observations towards the same source. The red-shifted filling emission feature appearing in another two independent OH/IR sources also confirms its reality. Even more strikingly, the spectral profiles of all `{\em RaBe}' type spectra associated with three different OH/IR sources look quite similar to each other not only in the blue-shifted filling emission feature but also in the whole wavelength range (from 52 to 55\,$\mu$m in Fig.\,\ref{groupsp1}).

OH/IR sources can also be classified according to the type of 53.3\,$\mu$m spectra. Generally, different spectra associated with the same OH/IR source should belong to the same type. This is true for almost all group-A OH/IR sources except \astrobj{IRAS 17574-2403} whose only two spectra belong to group `{\em BaRe}' and `{\em BroPf}' respectively. But the `{\em BroPf}' type spectrum looks strange if compared with the other spectrum of this source and its up and down scans show different line profiles. Therefore, this spectrum is unreliable. As a result, the 10 group A OH/IR sources together with the two megamasers can be classified into 3 `{\em RaBe}' type, 3 `{\em BaRe}' type, 2 `{\em Asymm}' type  and 4 `{\em GCMega}' type sources (Table\,\ref{prof53n35}). The profile of the 5 group T sources are also classified in the same way, but their classification is unreliable. 

How is the filling emission produced in the `{\em RaBe}' and `{\em BaRe}' type sources? One may argue that it can arise from some kind of emitting OH clumps with moving speed different from the absorbing OH clouds. But the amount of red-shift or blue-shift of these filling emission features from the line center (53.30\,$\mu$m) is measured to be about 0.1\,$\mu$m which is equivalent to a velocity shift of about 560\,km/s. The amount of red-shift and blue-shift are approximately the same in all spectra. Such a high velocity shift is impossible to explain from the expansion of the envelope around AGB stars, RSGs or H\,II regions. A more reasonable explanation is that the 53.3\,$\mu$m absorption is a doublet and one component of the doublet is in full absorption while the other is in partial absorption or even in emission. The theoretical separation of the two components can be found in ISAP line list database to be about 0.09\,$\mu$m, hence the distance from the two components to 53.30\,$\mu$m is about 0.45\,$\mu$m. Combined with the ISO instrument profile, the limited spectral resolution, the noise of the data and the expansion of the circumstellar envelope, this value basically agrees with the red-shift and blue-shift of the filling emission. But it is not clear how such asymmetric doublet can be produced in the OH maser shell.

The line profile classification is correlated with the IR object class mainly given by SIMBAD and included in Table\,\ref{prof53n35}. (The table will be further discussed in next section.) As it can be seen from the table, among the three `{\em RaBe}' type sources, \astrobj{IRAS 07209-2540} and \astrobj{NML\,Cyg} are well known red supergiants (RSGs) while \astrobj{IRAS 03507+1115} is a well studied nearby low mass evolved star. Most possibly this is an AGB star because it also shows SiO (Jewell et al.,\,\cite{jew91}) and H$_{2}$O (Bains et al.,\,\cite{bai03}) masers, and regular near infrared variation with a period of 470 days. Therefore this three sources can be considered as a group of evolved stellar OH/IR sources. \astrobj{IRAS 03507+1115} might be the first AGB star with its 53.3\,$\mu$m OH absorption line directly detected while its 34.6\,$\mu$m ISO SWS spectra are too noisy to show OH absorption. All three `{\em BaRe}' type sources: \astrobj{06053-0622},\astrobj{ 17574-2403} and \astrobj{20255+3712} are associated with known H\,II regions. Among the two `{\em Asymm}' type sources, one is associated to a H\,II region while the other is a stellar cluster which might be accompanied by H\,II regions too. The `{\em Asymm}' type spectra can also be considered as `{\em BaRe}' type spectra with weaker filling emisson at the red side of the line profile. Therefore the `{\em BaRe}' and `{\em Asymm}' type sources can be merged into one group: red-shifted filling emission type and hence the five OH/IR sources can be considered as a group of H\,II regions. Almost all spectra associated with H\,II regions are found to show strong ionized [OIII] emission at about 51.8$\mu$m, this confirms the H\,II nature of these IR sources. The ISO IR Spectral Energy Distribution (SED) peak position given in Table\,\ref{LineOHpump} also supports the spectral profile type--object class correlation, i.e., the SED of all `{\em RaBe}' type sources peaks at about 20\,$\mu$m (as evolved stellar OH/IR sources do) while that of many `{\em BaRe}' or `{\em Asymm}' type sources peaks at about 45\,$\mu$m (rather colder than the former).

In summary, evolved stellar OH/IR sources have `{\em RaBe}' type spectra showing strong blue-shifted filling emission in the 53.3\,$\mu$m line profile; H\,II regions have `{\em BaRe}' or `{\em Asymm}' type spectra showing red-shifted filling emission in the line profile; galaxy scale IR sources have `{\em GCMega}' type spectra mainly showing a deep absorption line with symmetrical profile precisely centered at 53.3\,$\mu$m. If the filling emission can be explained as filling emission in one of the 53.3\,$\mu$m doublet components, then one can conclude that evolved stellar OH/IR sources tend to show a filling emission in the blue component of the 53.3\,$\mu$m doublet, H\,II regions tend to show that in the red component while no prominent filling emission occurs in galaxy related sources. This difference may reflect the different physical conditions in the three kinds of different astronomical environments.
   \begin{figure}[]
   \centering
	 \includegraphics[width=9cm]{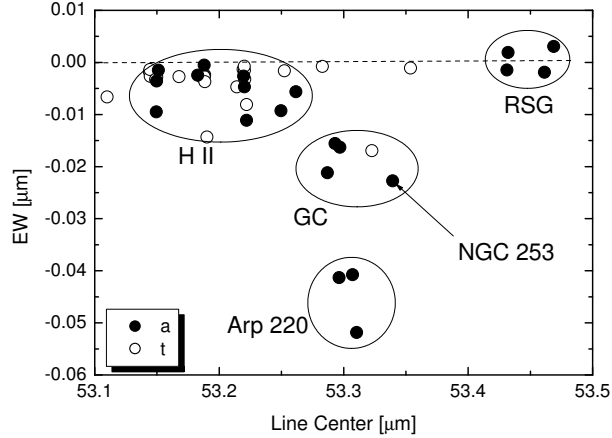}
   \caption{Equivalent width (EW) of 53.3\,$\mu$m absorption line shown against line center. Negative EW means absorption. Positive EW of some RSGs is due to strong filling emission. The spectra associated with different IR sources are distributed in separate areas. The arrow denotes the position of \astrobj{NGC\,253}.}
   \label{lc2-ew2}
   \end{figure}

The above correlation between spectral classification and object type can be further confirmed by the investigation of the 53.3\,$\mu$m absorption line center. The integrated line flux, center position and equivalent width of the 53.3\,$\mu$m line ($EW_{53.3}$) have been estimated for each group-a and t spectrum, but we only give a mean value of the spectral quantities for each OH/IR sources in Table\,\ref{LineOHpump}. The relation between $EW_{53.3} $ and line center is shown in Fig.\,\ref{lc2-ew2}. (The other quantities in the table will be discussed in next sections.) Negative $EW_{53.3}$ means absorption while positive $EW_{53.3}$ means emission. In this figure, spectra associated with different classes of objects, H\,II regions, evolved stellar OH/IR sources, GC sources and megamasers, gather in different areas. H\,II regions are located just at the blue side of 53.3\,$\mu$m due to the red-shifted filling emission. Evolved stellar OH/IR sources are located at the red side of 53.3\,$\mu$m due to the blue-shifted filling emission. GC sources and megamasers are located at just around 53.3\,$\mu$m. The $EW_{53.3}$ of the magamaser Arp\,220 is much larger than that of other sources. The $EW_{53.3}$ of GC sources and \astrobj{NGC\,253} is also larger than that of H\,II regions and evolved stellar OH/IR sources. The $EW_{53.3}$ of some evolved stellar OH/IR sources are positive due to the strong filling emission. This figure confirms the correlation between the line profile classification and the object class given in the upper discussion. The very large $EW_{53.3}$ of \astrobj{Arp\,220} indicates that the OH column density in this starburst galaxy is much larger than that in the Galaxy.

\section{Correlation between 34.6 and 53.3\,$\mu$m absorption lines}
\label{comp35and53}

The 34.6 and 53.3\,$\mu$m photons are generally considered as the main infrared pumping agency of radiatively pumped OH 1612\,MHz maser, and hence the absorption lines at this two wavelengths are expected to appear together in one and the same OH/IR source. The OH/IR sources with the line features detected or tentatively detected or with good quality spectra at both wavelengths are summarized in Table\,\ref{prof53n35}. The columns in the table are: (1)--source name; (2)--object class; (3)--IRAS/LRS spectral type; (4)--detection or spectral profile type of the 53.3\,$\mu$m line; (5)--detection of the 34.6\,$\mu$m line and (6)--the peak of spectral energy distribution (SED) estimated by observing the stamp plots of ISO SWS and/or LWS spectra by eyes. From this table, it can be seen that RSGs, GC sources and megamaser galaxies tend to show both 34.6 and 53.3\,$\mu$m absorptions; H\,II regions tend to show absorption at 53.3\,$\mu$m while their 34.6\,$\mu$m lines are not seen due to bad quality data; another two stellar sources, including one AGB star and one Wolf-Rayet star, show neither lines. 
\begin{table}[]
\caption[]{Summary of the 34.6 and 53.3\,$\mu$m absorption features of OH/IR sources and 
megamasers. Short solid lines are used to separate groups of different object types. Only 
those objects with the detection of the 34.6 or 53.3\,$\mu$m line or with good quality 
spectral data at both 34.6 and 53.3\,$\mu$m are listed in this table.}
\label{prof53n35}
\begin{tabular}{l@{ }l@{  }l@{  }l@{  }l@{  }l@{  }}
\hline
\noalign{\smallskip}
name            &obj.class      &LRS    &A53.3          &A34.6          &Peak [$\mu$m]   \\
     (1)                &(2)    &(3)    &(4)            &(5)            &(6)    \\
\noalign{\smallskip}                                                            
\hline                                                                          
\noalign{\smallskip}                                                            
$\astrobj{03507+1115}$    &AGB            &E      &RaBe           &N              &$<$30  \\
$\astrobj{07209-2540}$    &RSG            &E      &RaBe           &A              &19     \\
$\astrobj{NML\,Cyg}  $    &RSG            &$--$   &RaBe           &A              &19     \\
 --------       &               &       &               &               &       \\
$\astrobj{06053-0622}$    &H\,II          &H      &BaRe           &T              &44.5   \\
$\astrobj{17574-2403}$    &H\,II          &H      &BaRe           &N              &45     \\
$\astrobj{20255+3712}$    &H\,II          &H      &BaRe           &N              &$<$45  \\
$\astrobj{17424-2852}$    &H\,II          &I      &Asymm          &$--$           &45     \\
$\astrobj{17430-2848}$    &cluster        &H      &Asymm          &N              &40     \\
$\astrobj{17431-2846}$    &IR             &H      &Asymm(t)       &N              &45     \\
$\astrobj{18348-0526}$    &Mira           &A      &Asymm(t)       &U              &27     \\
$\astrobj{19244+1115}$    &pAGB           &E      &Asymm(t)       &A              &22     \\
$\astrobj{22176+6303}$    &H\,II          &H      &U1             &N              &45     \\
 --------       &               &       &               &               &       \\
$\astrobj{06319+0415}$    &IR             &H      &GCmega(t)      &$--$           &$<$45  \\
$\astrobj{16342-3814}$    &pAGB           &H      &GCmega(t)      &N              &45     \\
$\astrobj{17424-2859}$    &Sgr\,A*        &H      &GCmega         &A              &45     \\
$\astrobj{17441-2822}$    &Sgr\,B2        &$--$   &GCmega         &A              &80     \\
$\astrobj{Arp\,220}  $    &megamaser      &$--$   &GCmega         &A              &50     \\
$\astrobj{NGC\,253}  $    &megamaser      &P      &blue$-$tail    &N              &50     \\
 --------       &               &       &               &               &       \\
$\astrobj{17411-3154}$    &OH/IR          &A      &U1             &U              &35     \\
$\astrobj{10197-5750}$    &WR*            &H      &U2             &U              &28     \\
            \noalign{\smallskip}
            \hline
         \end{tabular}
\vspace*{1cm}
\end{table}
\begin{table}
\begin{list}{}{}
\scriptsize
\item[`--':] means no available data
\item[Column 2:]Object types from SIMBAD except several coming from literatures: \\
                \astrobj{03507+1115} (=IK TAU): We identify the star to be an AGB star from the maser quantities from 
                Jewell et al.~(\cite{jew91}) and Bains et al.~(\cite{bai03}) and some 
                well know facts;\\
                \astrobj{07209-2540} (= \astrobj{VY CMa}): Neufeld et al.~\cite{neu99} took it as a RSG; \\
                \astrobj{17424-2852}: H\,II region cross identified by Chen et al.~(\cite{che95});\\
                \astrobj{NML Cyg}: Justtanont et al.~\cite{jus96} took it as a RSG.
\item[Column 4:]`RaBe', `BaRe', `Asymm', `GCMega' are the 53.3\,$\mu$m spectral profile types while `U1' and `U2' are the 
                53.3\,$\mu$m data quality groups defined in Table~\ref{sptable}. `t' in a parenthesis means tentative detection and hence 
                the profile classifiction is not sure in this case. 
                Spectra of \astrobj{NGC\,253} belong to `GCMega' but show strong and very broad blue 
                absorption tail.
\item[Column 5:]Symbols: `A' and `T' mean the detection of the 34.6\,$\mu$m absorption is sure or tentative 
respectively, `U' means the expected 34.6\,$\mu$m line is not detected while `N' means the spectra are 
too noisy.
\item[Column 6:] The SED peak position is approximately estimated by observing the stamp 
plots of ISO SWS and/or LWS spectra by eyes.

\end{list}
\end{table}

\begin{figure}[]
   \centering
	 \includegraphics[width=9cm]{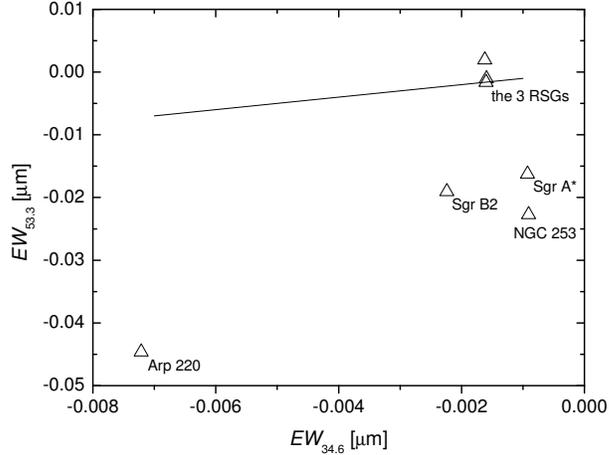}
   \caption{Mean equivalent width of 53.3\,$\mu$m absorption line $EW_{53.3}$ shown against that of 34.6\,$\mu$m absorption line ($EW_{34.6}$). The solid line represents equal EW of the two lines.}
   \label{EW53n35}
   \end{figure}

The two pumping lines can be further quantitatively compared by plotting the equivalent width of the 34.6\,$\mu$m lines ($EW_{34.6}$ from paper I) against that of the 53.3\,$\mu$m lines ($EW_{53.3}$ from Table\,\ref{LineOHpump}) for the five Galactic OH/IR sources and the two megamasers \astrobj{Arp\,220} and \astrobj{NGC\,253} of which both pumping lines are at least tentatively detected (Fig.\,\ref{EW53n35}) . As seen from this figure, the 53.3 to 34.6\,$\mu$m EW ratio is roughly equal to unity (near to the solid line which shows where the two EWs equal to each other) for RSGs but quite larger than unity for the four galaxy scale sources: \astrobj{Sgr\,A*} ($\approx 18$), \astrobj{Sgr\,B2} ($\approx 8$), \astrobj{Arp\,220} ($\approx 6$) and \astrobj{NGC\,253} ($\approx 25$). Both 34.6 and 53.3\,$\mu$m EWs are much larger for \astrobj{Arp 220} than for galactic sources and \astrobj{NGC\,253}. Although H\,II regions are not plotted in Fig.\,\ref{EW53n35} due to the lack of 34.6\,$\mu$m line data, one can take non--detections as detections with very weak line strength (below the 3 sigma noise level of the 34.6\,$\mu$m continuum flux). Therefore their 53.3 to 34.6\,$\mu$m EW ratio can be anticipated to be larger than that of RSGs, although their $EW_{53.3}$ is similar as or only slightly larger than that of the latter. The large 53.3 to 34.6\,$\mu$m EW ratio of the galaxy scale sources may be caused by interstellar OH absorption in front of a colder infrared emitting background. The large EW ratio of H\,II regions might be explained in a similar way. The fact that all three RSGs have their EW ratio close to unity indicates that equal EW of the two lines might be a general characteristic of isolated stellar OH 1612\,MHz masers while the much larger ratio might be a general characteristic of interstellar OH masers. The much larger EW of both lines in \astrobj{Arp\,220} may be the result of heavy interstellar OH absorption, because \astrobj{Arp\,220} is a starburst galaxy and huge volumns of the OH molecular clouds can exist in star forming regions.

\section{OH maser pumping}
\label{OHpump}

It has been shown in Sect.\,\ref{statonsp} that the 53.3\,$\mu$m line is detected or tentatively detected in $74\%$ of our good quality LWS spectra and in $33\%$ of the OH/IR sources with good quality spectra while it is not detected in $26\%$ of good quality spectra and in $16\%$ of OH/IR sources with good quality spectra. It is important to discuss whether the detections of the line support the radiative pumping mechanism and whether the non-detections undermine it. For the non-detections, the most natural explanation is that the OH maser may be genuinely very weak and hence the pumping line is too weak to be detected by ISO LWS spectrometer. But one can see from Fig.\,\ref{ISOcSOH} that some non--detections (several group U1, U2 and lots of U3, U4 sources) also have strong maser emission, therefore the  two OH pumping absorption lines should be strong too. Maybe the mechanisms discussed in Paper\,I, such as clumpy OH shell and limb filling emission, are still valid to explain the 53.3\,$\mu$m non--detections, otherwise these OH 1612\,MHz masers should not be radiatively pumped.

For the detection sources (of group A or T), the spectral profile classification results may give some hints on the pumping mechanism. All galactic stellar 53.3\,$\mu$m detections turn out to show filling emission in one of their hyperfine component, as discussed in Sect.\,\ref{profiles}. This may indicate that the asymmetric doublet profile or filling emission is a common phenomenon for galactic OH 1612\,MHz maser. Such an asymmetric filling emission might be best explained by the selective cascade of photons from energy levels higher than $^{2}\Pi _{1/2} (J=3/2)$. We also checked the 34.6\,$\mu$m absorption line profile of the three RSGs and found that the two components of the doublet are basically of equal strength in their SWS 07 spectra, therefore such an asymmetric 53.3\,$\mu$m doublet is highly probably produced by transitions among energy levels not higher than $^{2}\Pi _{1/2} (J=5/2)$. That is to say, transition from $^{2}\Pi _{1/2} (J=5/2)$ to $^{2}\Pi _{1/2} (J=3/2)$ in some kind of conditions can produce asymmetry in component strengths in the 53.3\,$\mu$m doublet but not in the 34.6\,$\mu$m doublet. The fact that the filling emission occurs in different components for RSGs and H\,II regions reflects the different physical conditions. If this is true, such asymmetric doublet profile may provide a strong proof to identify the radiative pumping of the OH 1612\,MHz masers from other pumping mechanisms. Further more, this filling emission can be considered also as a potential explanation for those non--detections of the line at 34.6 or 53.3\,$\mu$m or both.

Combining the 34.6 and 53.3\,$\mu$m data, one can go deeper into the maser pumping mechanism. The integrated 53.3\,$\mu$m line flux ($F^{int}_{53.3}$) and equivalent width (EW$_{53.3}$) are estimated by using their best quality spectra and the results are given in Table\,\ref{LineOHpump} (columns (2) and (3) respectively). Also given in the table are: column (4)--the integrated 34.6\,$\mu$m line flux ($F^{int}_{34.6}$)  from Paper I (the 34.6\,$\mu$m data of \astrobj{NGC\,253} is estimated in this paper from its SWS 02 spectrum TDT\,37902123); column (5)--integrated flux (blue peak + red peak) of the OH 1612\,MHz maser emission ($F^{int}_{OH}$)  obtained from literature (column (7)) and derived OH maser pump rates. The pump rate of radiatively pumped OH maser has been defined in Paper I as the integrated OH maser photon flux devided by the integrated 34.6\,$\mu$m IR absorption photon flux. Here we can define a better pump rate by deviding the integrated OH maser photon flux by the sum of the integrated absorption photon flux of  both pumping lines at 34.6 and 53.3\,$\mu$m (in column (6)).

It is easy to see from Table\,\ref{LineOHpump} that the pump rate of the three well known RSGs: \astrobj{07209-2540}, \astrobj{19244+1115} and \astrobj{NML Cyg}, are approximately the same. The pump rate of \astrobj{19244+1115} given here is a little higher than estimated using the photon rate data of Sylvester et al.~(\cite{syl97}), because our 34.6\,$\mu$m integrated flux is $69\%$ larger than theirs. This difference comes from different spectra used. (Readers are referred to paper I for detailed comparison.) The nearly similar pump rate of the three RSG OH masers suggests that their maser pumping mechanisms may be the same -- very possibly pure radiative pumping.

The pump rate calculated for the three galaxy scale sources show extreme values. The most prominent is the very small `pump rate' ($\sim 10^{-5}$ ) of the two GC sources: \astrobj{17424-2859} (=\astrobj{Sgr A*}) and \astrobj{17441-2822} (=\astrobj{Sgr B2}). Such a small pump rate supports the previously discussed large interstellar contribution to the strong 34.6 and 53.3\,$\mu$m absorption lines. The pump rate of the star burst galaxy, \astrobj{Arp 220}, is 0.32 which is much larger than that of RSGs and even larger than the theoretically predicted pump rate of 0.25. Such a large pump rate indicates that other pumping processes besides the radiative pumping, such as collisional pumping by shock waves,  must largely contribute to the observed megamaser. Lets take the RSGs as a standard template for radiatively pumped maser and designate it a mean pump rate of 0.06, then the contribution of radiative pumping to the megamaser of \astrobj{Arp\,220} is not higher than $0.06/0.32 \approx 19\%$. This is only an upper limit because, as discussed above, the two IR absorption lines are also due to interstellar OH clouds.
\begin{table*}[]{}
\caption[]{Mean line parameters of each group-A and T sources and other quantities 
for OH maser radiative pump rate estimation. Fluxes $F^{int}_{53.3}$, 
$F^{int}_{34.6}$ are in unit of [$10^{-19}$W/cm$^2$], $F^{int}_{OH}$ is in unit of 
[$10^{-26}$W/cm$^2$], equivalent width EW is in unit of [$10^{-4}\mu$m]. $F^{int}_{OH}$ 
is expressed by (blue component + red component). Negative flux means absorption line.}
\label{LineOHpump}
\begin{tabular}{l@{ }r@{  }r@{  }r@{  }c@{  }l@{  }l@{  }}
\hline
\noalign{\smallskip}
name            &$F^{int}_{53.3}$       &EW$_{53.3}$    &$F^{int}_{34.6}$       &$F^{int}_{OH}$ &pumprate       &notes$^*$      \\
(1)             &(2)                    &(3)            &(4)                    &(5)            &(6)            &(7)            \\
\noalign{\smallskip}                                                            
\hline                                                                          
\noalign{\smallskip}                                                            
\multicolumn{7}{l}{\textbf{group A}}                            \\
$\astrobj{03507+1115}$    &$1.37   $        &$30.8   $        &$     $          &$           $    &$       $        &               \\
$\astrobj{06053-0622}$    &$-25.33 $        &$-14.7  $        &$     $          &$           $    &$       $        &               \\
$\astrobj{07209-2540}$    &$3.96   $        &$19.4   $        &$-17.6$          &$(970+781)  $    &$0.079  $        &tL-H(fig)      \\
$\astrobj{17424-2852}$    &$-50.34 $        &$-92.5  $        &$     $          &$           $    &$       $        &               \\
$\astrobj{17424-2859}$    &$-314.79$        &$-162.6 $        &$-27.7$          &$(1.98+1.53)$    &$3.56e-6$        &LWH92          \\
$\astrobj{17430-2848}$    &$-21.72 $        &$-111.1 $        &$     $          &$           $    &$       $        &               \\
$\astrobj{17441-2822}$    &$-129.3 $        &$-190.5 $        &$-1.10$          &$(3.15+2.85)$    &$1.56e-5$        &ATCAb(fig)     \\
$\astrobj{17574-2403}$    &$-93.2  $        &$-56.2  $        &$     $          &$           $    &$       $        &               \\
$\astrobj{20255+3712}$    &$-26.52 $        &$-28.8  $        &$     $          &$           $    &$       $        &               \\
\astrobj{Arp\,220}        &$-5.36  $        &$-446.3 $        &$-1.00$          &$5667^{**}$      &$0.32   $        &baa87(fig)     \\
\astrobj{NGC\,253}        &$-23.23 $        &$-227.4 $        &$-0.72^{\#}$          &$           $    &$       $        &               \\
\astrobj{NML\,Cyg}        &$-2.08  $        &$-16.5  $        &$-11.5$          &$(1130+24)  $    &$0.041  $        &priv(fig)      \\
\noalign{\smallskip}                                                    
\multicolumn{7}{l}{\textbf{group T}}                            \\
\noalign{\smallskip}                                                    
$\astrobj{06319+0415}$    &$-0.83$          &$-7.68$          &$     $          &$   $            &$     $          &               \\
$\astrobj{16342-3814}$    &$-5.39$          &$-143 $          &$     $          &$   $            &$     $          &               \\
$\astrobj{17431-2846}$    &$-31.6$          &$-80.6$          &$     $          &$   $            &$     $          &               \\
$\astrobj{18348-0526}$    &$-2.82$          &$-66.2$          &$     $          &$   $            &$     $          &               \\
$\astrobj{19244+1115}$    &$-1.12$          &$-10.5$          &$-8.10$          &$662$            &$0.054$          &SYL97          \\
            \noalign{\smallskip}
            \hline
         \end{tabular}
\begin{list}{}{} 
\scriptsize      
\item[*:] The reference codes for OH 1612\,MHz maser data are the same as in Table\,\ref{objtable} except: 
	  SYL97: Sylvester et al.~(\cite{syl97})\\
          (fig) means the integrated 1612\,MHz OH maser fluxes are derived 
          by measuring from the figure published in the paper.\\
\item[**:] There is only a single broad maser emission feature for \astrobj{Arp\,220}.
\item[\#:] The 34.6\,$\mu$m line flux of \astrobj{NGC\,253} is estimated by this paper 
from its SWS\,02 spectrum TDT\,37902123. The 34.6\,$\mu$m line is only tentatively detected 
in this spectrum.
\end{list}
\end{table*}

\section{Summary}
\label{summary}

Based on the statistical analysis of the 102 ISO LWS spectra (out of 137) associated with 47 galacitic OH/IR sources and another 4 spectra of two megamasers, we discussed the detectability of the 53.3\,$\mu$m OH absorption, line profile classification and OH maser pumping. The main conclusions can be summarized as follow:

\begin{enumerate}
\item The 106 ISO LWS spectra associated with 47 galactic OH/IR sources and two megamasers are grouped according to the detection of the 53.3\,$\mu$m line and 3 sigma noise level into 7 groups: a, t, u1, u2, u3, u4 and u5. The detection and non--detection rate of the line based on the spectra counting and free of the instrumental sensitivity and resolution selection effect is about $74\%$ (32 spectra out of 43) and $26\%$ (11 spectra out of 43) respectively.
\item The 49 OH/IR sources are also grouped according to the detection of the 53.3\,$\mu$m line in their spectra into 7 groups: A, T, U1, U2, U3, U4 and U5. The detection and non--detection rate of the line based on the galactic sources counting is about $33\%$ (15 sources out of 45) and $16\%$ (7 sources out of 45) respectively.
\item The blue peak flux of the 1612\,MHz OH maser is found to decrease with increasing ISO infrared color $C_{ISO}$ and this trend is mainly supported by H\,II regions (with IRAS/LRS type `H').
\item The 53.3\,$\mu$m line profiles can be classified into 6 types: {\em RaBe}, {\em BaRe}, {\em Asymm}, {\em BroPf}, {\em SpurEm}, {\em GCMega}. The filling emission appearing in the first three types of spectra is argued to be  the manifestation of  the asymmetric 53.3\,$\mu$m doublet with one of its find structure components in fully absorption while the other partially in absorption or even in emission.
\item The OH/IR sources with 53.3\,$\mu$m detection are also classified according to the line profile into 4 classes: {\em RaBe}, {\em BaRe}, {\em Asymm}, {\em GCMega}. The {\em RaBe} type sources are all evolved stellar OH/IR sources (2 RSGs and 1 AGB star) while the {\em BaRe} and {\em Asymm} type sources (3 sources and 2 sources respectively) are usually H\,II regions. That is to say, evolved stellar OH/IR sources tend to show filling emission in the blue doublet component of their 53.3\,$\mu$m line, H\,II regions tend to show that in the red doublet component of the line while galaxy scale sources do not show such asymmetry in the line profile. \astrobj{IRAS 03507+1115} is the first AGB star directly found to show the 53.3\,$\mu$m pumping line.
\item GC sources show large 53.3\,$\mu$m line equivalent width ($EW_{53.3}$) but small 34.6\,$\mu$m line equivalent width ($EW_{34.6}$); RSGs have small $EW_{34.6}$ and $EW_{53.3}$; H\,II regions also have small $EW_{53.3}$ but do not show 34.6\,$\mu$m absorption; the megamaser \astrobj{Arp\,220} has its $EW_{34.6}$ and $EW_{53.3}$ both very large. The large $EW_{53.3}$ of all four galaxy scale sources and large $EW_{34.6}$ of Arp\,220 may be all due to interstellar OH absorption. The $EW_{53.3}$ to $EW_{34.6}$ ratio is close to unity for RSGs but quite larger than unity for the other three kinds of objects: H\,II regions, GC sources and megamaser, which may also reflect a larger interstellar contribution.
\item If the OH 1612\,MHz masers in some RSGs and H\,II regions of our sample are really radiatively pumped, the filling emission in one of the 53.3\,$\mu$m doublet components is probably produced by the transitions between the two OH energy levels: $^{2}\Pi _{1/2} (J=3/2)$ and $^{2}\Pi _{1/2} (J=5/2)$ or their closely related levels. The fact that the filling emission occurs in different components for the two kinds of objects may reflect different physical conditions.
\item The pump rate of RSGs is about 0.06 while that of GC sources is very small and that of \astrobj{Arp\,220} is very large. Perhaps the pump rate of RSGs represents a typical value for purely radiatively pumped OH 1612\,MHz masers. Low pump rate of GC sources is due to interstellar OH aborption while very large pump rate of Arp\,220 indicates a large contribution of other pumping mechanisms (e.g., collisional pumping).
\item A handful of OH/IR sources with both pumping lines not detected in their better quality spectra still need to be explained. The genuinely weakness of the OH maser and hence of the two IR pumping lines can explain part of these non--detections. Maybe the alternative explanations presented in Paper I, such as clumpy OH shell and limb filling emission, are also valid.
\end{enumerate}

\ack{Guest User, Canadian Astronomy Data Center, which is operated by the Dominion Astrophysical Observatory for the National Resarch Council of Canada's Herzberg Institute of Astrophysics. We thank Dr. R. Szczerba for the help in processing ISO spectra and those astronomers who have prepared the ISO projects to observe the OH/IR sources used in this paper. Our work is supported by  the Chinese National Science Foundation under Grant No. 10073018, the Chinese Academy of Sciences Foundation under Grant KJCX2-SW-T06 and the Yunnan Natural Science Fund (2002A0021Q).
}

\end{document}